\documentclass[twocolumn,prd,nofootinbib]{revtex4}
\usepackage{amsmath}
\usepackage{graphicx}

\begin{document}

\title{No Swiss-cheese universe on the brane}
\author{L\'{a}szl\'{o} \'{A}. Gergely\thanks{
gergely@physx.u-szeged.hu}}
\affiliation{Departments of Theoretical and Experimental Physics, University of Szeged,
Szeged 6720, D\'{o}m t\'{e}r 9, Hungary}

\begin{abstract}
We study the possibility of brane-world generalization of the
Einstein-Straus Swiss-cheese cosmological model. We find the modifications
induced by the brane-world scenario. At a first glance only the motion of
the boundary is modified and the fluid in the exterior region is allowed to
have pressure. The general relativistic Einstein-Straus model emerges in the
low density limit. By imposing that the brane is static, a combination of
the junction conditions and modified cosmological evolution leads to the
conclusion that the brane is flat. Thus no static Swiss-cheese universe can
exist on the brane. The conclusion is not altered by the introduction of a
cosmological constant in the FLRW regions. This result mimics a similar
general relativistic result: static Einstein-Straus universes do not exist.
\end{abstract}

\date{\today }
\startpage{1}
\endpage{}
\maketitle

\section{Introduction}

Brane-world scenarios introduced by Arkani-Hamed, Dimopoulos and Dvali \cite%
{ADD} and Randall and Sundrum (RS) \cite{RS2} are motivated by string
theory, where open strings end on branes. In the generalized RS scenario our
4-dimensional Friedmann-Lema\^{\i}tre-Robertson-Walker (FLRW) universe is a
hypersurface of codimension one (the brane) embedded into a 5-dimensional
charged Vaidya-Anti de Sitter or Reissner-Nordstr\"{o}m-Anti de Sitter
space-time (the bulk). The radiation in the bulk, the charge and mass of the
central bulk black hole can be switched off independently (for a detailed
discussion see \cite{Decomp}), however the bulk will always have a negative
cosmological constant, which balances the brane tension in such a way that
the 4-dimensional (4D) cosmological constant can be fine-tuned. Branes with
various other symmetries were also examined, like the Einstein brane \cite%
{Einbrane}, a Kantowski-Sachs type brane \cite{Einbrane2} and the G\"{o}del
brane \cite{BTs}. All these cases bear in common unusual properties or
exotic matter.

There were also various attemps to find black hole solutions on the brane
(for a recent review see \cite{MaartensLR}). For example in \cite{DMPR} the
charge term of the 4D Reissner-Nordstr\"{o}m black hole solution was
interpreted as arising from a tidal charge of the bulk Weyl tensor (and
having arbitrary sign), rather than from electric charge. The bulk
containing this black hole however remains unknown. Gravitational lensing 
\cite{Whisker} and galactic rotation curves \cite{MakHarko} were also
studied in brane-world models.

Black holes are not isolated objects. In any realistic model they should be
embedded in cosmological background. The simplest \textit{general
relativistic} cosmological scenario admitting local inhomogeneities is the
Einstein-Straus Swiss-cheese model \cite{ES}, in which Schwarzschild black
holes and surrounding vacuum regions are immersed into a FLRW background.
This model was employed to show that cosmic expansion has no influence on
planetary orbits and also to give corrections to the luminosity-redshift
relation \cite{Kantowski}. The Einstein-Straus model however is unstable
against perturbations. (It is also unsuitable for other type of symmetry
then spherical, like cylindrical \cite{Senovilla} or axial \cite{Mars}.) A
more realistic model suitable for encompassing (spherically symmetric)
inhomogeneities would be the McVittie solution \cite{McVittie}, which allows
for both the FLRW and the Schwarzschild limits. According to this model, the
cosmic expansion drives the planetary orbits into an outward spiraling and
galactic clusters into expansion, which however happen more slowly than the
expansion of the cosmic background and consequently these effects are
undetectable. An excellent review of these topics can be found in the book
of Krasi\'{n}ski \cite{Krasinski}. Recently, the McVittie solution was
generalized both to arbitrary dimensions \cite{Gao} and to include charge in
4D \cite{GaoZhang}.

\begin{figure}[tbp]
\vskip-0.5cm 
\includegraphics[height=8cm]{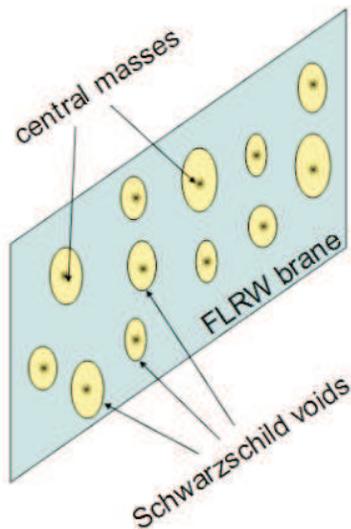} 
\caption{(Color online.) A Swiss-cheese brane-world.}
\label{Fig1}
\end{figure}

\begin{figure}[tbp]
\vskip-1cm 
\centering\includegraphics[width=6.3cm]{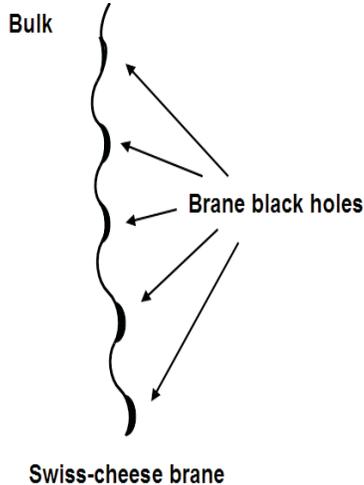} 
\caption{A qualitative picture showing that the inhomogeneous extrinsic
curvature of an inhomogeneous brane leads to humps and bumps of the brane
embedding in the bulk.}
\label{Fig2}
\end{figure}

Such cosmological models with local inhomogeneities were not studied before
in the context of the generalized RS scenario. It is the aim of this paper
to study the simplest such model, the Swiss-cheese on the brane. The
difficulty in such an approach is threefold. First, one has to join
(according to the Lanczos-Sen-Darmois-Israel junction conditions \cite%
{Lanczos}-\cite{Israel}) black hole solutions with the cosmological
background on the brane, both being solutions of a \textit{modified}
Einstein equation \cite{SMS}, such that there is no distributional matter on
the junction surface (Fig. \ref{Fig1}). Second, one has to extend somehow
these black hole solutions into the bulk, which is far from trivial. Indeed,
the simplest 4D Schwarzschild solution could be embedded in the bulk only by
extending the singularity into the fifth dimension \cite{CHR}, obtaining a 
\textit{black string}. In order to obtain a Schwarzschild brane black hole
with regular AdS horizon, exotic matter has to be introduced in the bulk 
\cite{KantiTamvakis}. If the horizon is not compactified through the fifth
dimension, gravitons from the black hole will escape into the fifth
dimension even if there is an event horizon surrounding the hole on the
brane. The third task is to interpret \textit{geometrically} such an
inhomogeneous brane as a junction hypersurface. Obviously, for different
matter content $\tau _{ab}=-\lambda g_{ab}+T_{ab}$ in the voids and outside
them (where $\lambda $ is the brane tension, $g_{ab}$ the brane metric and $%
T_{ab}$ represents the energy-momentum tensor of ordinary matter on the
brane), the Lanczos equation 
\begin{equation}
\Delta K_{ab}=-\widetilde{\kappa }^{2}\left( \tau _{ab}-\frac{\tau }{3}%
g_{ab}\right) \   \label{Lanczos}
\end{equation}%
(with $\Delta K_{ab}=-2K_{ab}$ for $Z_{2}$-symmetric embedding) gives rise
to different extrinsic curvatures $K_{ab}$ for the voids and for the rest of
the brane. Although the embedding (given by the 1-form $n=dy$, the brane
being at $y=0$) and the extrinsic curvature $K_{ab}=g_{a}^{c}{}g_{b}^{d}%
\widetilde{\nabla }_{c}n_{d}$ are closely interrelated, the latter also
depends on the bulk metric $\widetilde{g}_{ab}=g_{ab}+n_{a}n_{b}$ and the
associated connection $\widetilde{\nabla }$. Some of the difference in $%
K_{ab}$ at voids and expanding background can be explained by the difference
in $\widetilde{g}_{ab}$ due to how near or far the voids are, however the
rest should be interpreted as humps and bumps in the brane embedding (Fig. %
\ref{Fig2}).

In the simplest Swiss-cheese model on the brane we will see that fulfilling
the first task is restrictive enough to make the second and third ones
redundant.

\section{A no-go result}

We suppose the bulk contains nothing but a cosmological constant and its
Weyl curvature is such that it has no electric part. The embedding is $Z_{2}$%
-symmetric. Under these assumptions the modified Einstein equation is 
\begin{equation}
G_{ab}=-\Lambda g_{ab}+\kappa ^{2}T_{ab}+\widetilde{\kappa }^{4}S_{ab}\ ,
\label{modE}
\end{equation}%
where $S_{ab}$ denotes a quadratic expression in $T_{ab}$: 
\begin{equation}
S_{ab}={{\frac{1}{12}}}TT_{ab}-{{\frac{1}{4}}}T_{ac}T^{c}{}_{b}+{{\frac{1}{24%
}}}g_{ab}\left( 3T_{cd}T^{cd}-T^{2}\right) \,.  \label{S}
\end{equation}%
The brane gravitational constant $\kappa ^{2}$ and the brane cosmological
constant\ $\Lambda $ are related to the bulk gravitational constant $%
\widetilde{\kappa }^{2}$, bulk cosmological constant $\widetilde{\Lambda }$
and the (positive) brane tension $\lambda $ through 
\begin{align}
6\kappa ^{2}& =\widetilde{\kappa }^{4}\lambda \ , \\
2\Lambda & =\kappa ^{2}\lambda +\widetilde{\kappa }^{2}\widetilde{\Lambda }\
.
\end{align}

The brane is a 4-dimensional FLRW space-time (characterized by scale factor $%
a\left( \tau \right) $). It contains perfect fluid (characterized by energy
density $\rho $ and pressure $p$, both depending only on cosmological time $%
\tau $). The inhomogeneities on the brane are introduced as nonintersecting
Schwarzschild voids with arbitrary radii. Let us study the junction of one
of these voids to the rest of the brane. The interior solution in curvature
coordinates $\left( T,R\right) $ is%
\begin{eqnarray}
ds_{S}^{2} &=&-\left( 1-\frac{2m}{R}\right) dT^{2}+\left( 1-\frac{2m}{R}%
\right) ^{-1}dR^{2}  \notag \\
&&+R^{2}\left( d\theta ^{2}+\sin ^{2}\theta d\varphi ^{2}\right) \ .
\label{Schw}
\end{eqnarray}%
This is a pure vacuum solution (with $\Lambda =0$) of the modified Einstein
equation (\ref{modE}) in the void regions. The exterior solution in comoving
coordinates is%
\begin{gather}
ds_{FLRW}^{2}=-d\tau ^{2}+a^{2}\left( \tau \right)  \notag \\
\times \left[ d\chi ^{2}+\mathcal{H}^{2}\left( \chi ;k\right) \left( d\theta
^{2}+\sin ^{2}\theta d\varphi ^{2}\right) \right] \ ,  \label{FLRW} \\
\mathcal{H}(\chi ;k)=\left\{ 
\begin{array}{cc}
\sin \ \chi \  & ,\qquad k=1\,, \\ 
\chi & ,\qquad k=0\,, \\ 
\sinh \ \chi & \ ,\qquad k=-1\,.%
\end{array}%
\right.  \notag
\end{gather}%
Here the function $\mathcal{H}$ has the properties:%
\begin{eqnarray}
\left( \frac{d\mathcal{H}}{d\chi }\right) ^{2} &=&1-k\mathcal{H}^{2}\ ,
\label{p1} \\
\frac{d^{2}\mathcal{H}}{d\chi ^{2}} &=&-k\mathcal{H}\ .  \label{p2}
\end{eqnarray}%
The modified Einstein equations (\ref{modE}) reduce to a generalized
Friedmann and a generalized Raychaudhuri equation:%
\begin{gather}
\frac{\dot{a}^{2}+k}{a^{2}}=\frac{\Lambda }{3}+\frac{\kappa ^{2}\rho }{3}%
\left( 1+\frac{\rho }{2\lambda }\right) \ ,  \label{Fried} \\
\frac{\ddot{a}}{a}=\frac{\Lambda }{3}-\frac{\kappa ^{2}}{6}\left[ \rho
\left( 1+\frac{2\rho }{\lambda }\right) +3p\left( 1+\frac{\rho }{\lambda }%
\right) \right] \ .  \label{Raych}
\end{gather}%
We have kept the cosmological constant in the exterior region for later
convenience. In the limit $\rho /\lambda \rightarrow 0$ the corresponding
general relativistic equations are recovered. We have identified the angular
coordinates $\theta $ and $\varphi $ in the two regions, as in principle the
comoving coordinate systems can be centered on any chosen Schwarzschild hole.

The junction conditions require the continuity of both the first and second
fundamental forms (induced metric and extrinsic curvature) of the junction
hypersurface. In the Swiss-cheese model the junction is made at some
constant comoving $\chi =\chi _{0}$. Therefore it is obvious to introduce $%
\left( \tau ,\theta ,\varphi \right) $ as coordinates on the junction
hypersurface. The induced metrics in the two regions become: 
\begin{eqnarray}
ds_{int}^{2} &=&\left[ -\left( 1-\frac{2m}{R_{0}}\right) \dot{T}%
_{0}^{2}+\left( 1-\frac{2m}{R_{0}}\right) ^{-1}\dot{R}_{0}^{2}\right] d\tau
^{2}  \notag \\
&&+R_{0}^{2}\left( d\theta ^{2}+\sin ^{2}\theta d\varphi ^{2}\right) \ , \\
ds_{ext}^{2} &=&-d\tau ^{2}+a^{2}\left( \tau \right) \left[ \mathcal{H}%
_{0}^{2}\left( d\theta ^{2}+\sin ^{2}\theta d\varphi ^{2}\right) \right] \ ,
\end{eqnarray}%
where $R_{0}=R\left( \tau ,\chi _{0}\right) ,\ T_{0}=T\left( \tau ,\chi
_{0}\right) $ and $\mathcal{H}_{0}=\mathcal{H}\left( \chi _{0};k\right) $.
Continuity of the induced metric implies%
\begin{eqnarray}
R_{0} &=&a\left( \tau \right) \mathcal{H}_{0}\ ,  \label{c1} \\
\left( 1-\frac{2m}{a\left( \tau \right) \mathcal{H}_{0}}\right)^{2}\!\! \dot{%
T}_{0}^{2} \!\! &=&\!\! 1-\frac{2m}{a\left( \tau \right) \mathcal{H}_{0}}+%
\dot{a}^{2}\left( \tau \right) \mathcal{H}_{0}^{2}\ .  \label{c2}
\end{eqnarray}%
These equations characterize the motion of the boundary region of the
Schwarzschild void and formally they are the same as in the Einstein-Straus
model. However due to the modified cosmological evolution Eqs. (\ref{Fried}%
)-(\ref{Raych}), \textit{the motion of the boundary is changed }accordingly.

The extrinsic curvature of the junction hypersurface evaluated from the
exterior region has only two nonvanishing components:%
\begin{eqnarray}
K_{\theta \theta }^{ext} &=&a\left( \tau \right) \mathcal{H}_{0}\left( 1-k%
\mathcal{H}_{0}^{2}\right) ^{1/2}\ ,  \notag \\
K_{\varphi \varphi }^{ext} &=&K_{\theta \theta }\sin ^{2}\theta \ .
\end{eqnarray}%
(We have employed Eq. (\ref{p1}) in the derivation). The corresponding
extrinsic curvature components evaluated from the interior region are: 
\begin{eqnarray}
K_{\theta \theta }^{int} &=&\left( 1-\frac{2m}{R_{0}}\right) R_{0}\dot{T}%
_{0}\ ,  \notag \\
K_{\varphi \varphi }^{int} &=&K_{\theta \theta }\sin ^{2}\theta \ .
\end{eqnarray}%
From continuity of these components, by employing Eq. (\ref{c1}) we obtain a
simple equation for $\dot{T}_{0}$ 
\begin{equation}
\dot{T}_{0}=\left( 1-\frac{2m}{a\left( \tau \right) \mathcal{H}_{0}}\right)
^{-1}\left( 1-k\mathcal{H}_{0}^{2}\right) ^{1/2}\ .  \label{c3orig}
\end{equation}%
Comparison of Eqs. (\ref{c2}) and (\ref{c3orig}) gives%
\begin{equation}
\dot{a}^{2}\left( \tau \right) +k=\frac{2m}{a\left( \tau \right) \mathcal{H}%
_{0}^{3}}\ .  \label{c3}
\end{equation}%
Finally the condition $K_{\tau \tau }^{int}=0$ implies%
\begin{equation}
\ddot{a}\left( \tau \right) =-\frac{m}{a^{2}\left( \tau \right) \mathcal{H}%
_{0}^{3}}\ ,  \label{c4}
\end{equation}%
after simplifying by $d\mathcal{H}/d\chi \mid _{\chi =\chi _{0}}$, which is
a nonvanishing constant. As the evolution of the scale factor is known (Eqs.
(\ref{Fried}) and (\ref{Raych})), we can express both the central mass and
the pressure in terms of the energy density:%
\begin{eqnarray}
m &=&\left[ \Lambda +\kappa ^{2}\rho \left( \tau \right) \left( 1+\frac{\rho
\left( \tau \right) }{2\lambda }\right) \right] \frac{a^{3}\left( \tau
\right) \mathcal{H}_{0}^{3}}{6}\ ,  \label{m} \\
p\left( \tau \right)  &=&\frac{\Lambda }{\kappa ^{2}\left( 1+\frac{\rho
\left( \tau \right) }{\lambda }\right) }-\frac{\rho \left( \tau \right) ^{2}%
}{2\left[ \rho \left( \tau \right) +\lambda \right] }\ .  \label{pp}
\end{eqnarray}%
Due to cosmological symmetries, these equations derived on the boundary are
valid everywhere in the FLRW brane. Therefore Eq. (\ref{m}) can be also
regarded as a relation between the mass and the comoving radius of the
Schwarzschild void.

There are two major differences in comparison to the general relativistic
model. The first is, that the cosmological perfect fluid is \textit{not}
dust. In the general relativistic limit $\kappa ^{2}p=\Lambda $. If the
cosmological constant is chosen to vanish, similarly as in the interior
region, the fluid is dust. By contrast, in the brane-world scenario the
nonlinear source term of Eq. (\ref{modE}) implies a $\tau $-dependent
pressure (or tension). This agrees with the general relativistic limit in
the low density regime ($\rho <<\lambda $), however differs considerably in
the high energy limit ($\rho >>\lambda $), where $p\simeq -\rho /2$. While
in this later limit the classical condition for dark energy $\rho +3p\simeq
-\rho /2<0$ is obeyed, still the perfect fluid with equation of state (\ref%
{pp}) cannot drive inflation. Indeed, the cosmic acceleration is given by
the \textit{modified} Raychaudhuri equation (\ref{Raych}), which becomes:%
\begin{equation}
\frac{\ddot{a}}{a}=-\frac{\Lambda }{6}-\frac{\kappa ^{2}}{6}\rho \left( 1+%
\frac{\rho }{2\lambda }\right) \ ,  \label{Raych1}
\end{equation}%
giving deceleration for any positive $\Lambda $.

The second crucial difference is the evolution of the fluid energy density, 
which differs from the general relativistic case. To see this, we integrate 
the continuity equation: 
\begin{equation}
\dot{\rho}+3\frac{\dot{a}}{a}\left( \rho +p\right) =0\   \label{conti}
\end{equation}%
(which is an integrability condition of the system of Eqs. (\ref{Fried})-(%
\ref{Raych})). After employing Eq. (\ref{pp}) we obtain by integration 
\begin{equation}
a^{3}=C\frac{\kappa ^{2}}{2\lambda }\left[ \Lambda +\kappa ^{2}\rho \left( 1+%
\frac{\rho }{2\lambda }\right) \right] ^{-1}\ ,  \label{rhoaa}
\end{equation}%
with $C$ an integration constant. Eq. (\ref{m}) gives then the mass of the
black holes 
\begin{equation}
m=\frac{\kappa ^{2}C}{12\lambda }\mathcal{H}_{0}^{3}\ .  \label{m2}
\end{equation}
in the voids of the Einstein-Straus
type Swiss-cheese branes. 

\textit{Case }$\Lambda =0$\textit{.} This is the standard choice in the
Swiss-cheese model. In the \textit{low density} limit $\rho \sim a^{-3}$
emerges, as in the general relativistic case. In the special case of \textit{%
static }branes ($a=a_{1}$ constant), the continuity Eq. (\ref{conti})
implies $\rho =\rho _{1}=$const and Eqs. (\ref{rhoaa}) and (\ref{m2}) do not
arise any more.$\ $Then Eq. (\ref{pp}) gives $p=p_{1}=$const as well.
Moreover Eq. (\ref{Raych1}) implies $\rho _{1}=0$ and then Eq. (\ref{pp})
gives $p_{1}=0$, thus the FLRW region must be empty. Further, Eq. (\ref{c4})
gives $m=0$ and then Eq. (\ref{c3}) $k=0$, thus the whole brane becomes trivial, a 
\textit{flat} hypersurface.  

\textit{Case }$\Lambda \neq 0$\textit{.} Again, for \textit{static }branes
it is evident from the continuity Eq. (\ref{conti}) that $\rho =\rho _{2}=$%
const should hold, while from Eq. (\ref{pp}) $p=p_{2}=$const emerges. The
Raychaudhuri equation (\ref{Raych1}) gives for the energy density of the
fluid: 
\begin{equation}
\rho _{2}=-\lambda \pm \sqrt{\lambda \left( \lambda -2\Lambda /\kappa
^{2}\right) }\   \label{rho1}
\end{equation}%
and also $\Lambda <0$. (Therefore for a positive energy density we have to
choose the $+$ sign in Eq. (\ref{rho1}).) From Eq. (\ref{pp}) after some
straightforward algebra we find the pressure:%
\begin{equation}
p_{2}=-\rho _{2}\ .  \label{pp1}
\end{equation}%
The cosmological fluid $T_{ab}=-\rho _{2}g_{ab}$ turns out to be just
another contribution to the cosmological constant. The central mass is
vanishing by virtue of Eq. (\ref{c4}) and then Eq. (\ref{c3}) implies once
more $k=0$. The brane is again \textit{flat,} the 4-dimensional Minkowski
space-time. Both a cosmological constant and a nonvanishing perfect fluid
are present on the FLRW region of the brane, and they do not cancel, however
the third, quadratic source term (\ref{S}) of the modified Einstein equation%
\begin{equation}
\widetilde{\kappa }^{4}S_{ab}=-{{\frac{\kappa ^{2}\rho _{2}^{2}}{2\lambda }}}%
g_{ab}\,
\end{equation}%
cancels both by virtue of Eq. (\ref{rho1}) 
\begin{eqnarray}
-\Lambda g_{ab}+\kappa ^{2}T_{ab}+\widetilde{\kappa }^{4}S_{ab} &=&  \notag
\\
-\kappa ^{2}\left[ \Lambda /\kappa ^{2}+\rho _{2}\left( 1+{{\frac{\rho _{2}}{%
2\lambda }}}\right) \right] g_{ab} &=&0\ .
\end{eqnarray}

\section{Concluding Remarks}

We have found that as in the general relativistic case (in the
Einstein-Straus model), the voids on a Swiss-cheese type brane-world reside
in a dynamical universe. The density of the fluid in the Swiss-cheese brane
evolves cf. Eq. (\ref{rhoaa}) in a different way as compared to general
relativity. A static Swiss-cheese universe cannot exist neither on the
brane, nor in general relativity, apart from the trivial empty and flat
universe. 

Our no-go theorem for static Swiss-cheese branes adds to various results
concerning the staticity of brane stellar models \cite{ND}-\cite{Casadio}.
In particular, in \cite{BGM} it was shown that for an Oppenheimer-Sneider
collapse on the brane the exterior space-time (on the brane) cannot be
static. Similarly it was shown that the vacuum exterior of a spherical star
(with vanishing pressure at the surface) is in general not Schwarzschild 
\cite{GM}. In \cite{ND} the condition of vanishing pressure at the junction
surface was lifted, as in our discussion.

It remains to see what type of inhomogeneous brane cosmological model can be
found by lifting some of the original assumptions of the Swiss-cheese model.
An obvious way to do this is by allowing a non-vanishing electric part of
the bulk Weyl curvature, which would contribute as source of the modified
Einstein equation. In such a scenario the voids would be non-vacuum
solutions in the classical general relativistic sense. Investigations of
these issues are under way.

\section{Acknowledgements}

I am grateful for references, discussions and comments on the manuscript to
Viktor Czinner, Nathalie Deruelle, Roy Maartens, M\'{a}ty\'{a}s Vas\'{u}th
and Zolt\'{a}n Perj\'{e}s, who was a gifted teacher and wonderful friend.
This work was supported by OTKA grants no. T046939 and TS044665.

\end{document}